\documentclass[manuscript)]{aastex}

\usepackage{lscape,graphicx,epsfig,natbib,color}
\usepackage{epstopdf}
\makeatletter

\newcommand{\Rmnum}[1]{\expandafter\@slowromancap\romannumeral #1@}
\makeatother

\slugcomment{Submitted to ApJ}

\shorttitle{Chromospheric condensation and QPPs in a circular-ribbon flare}
\shortauthors{Zhang et al.}

\begin{document}
\title{Chromospheric Condensation and Quasi-periodic Pulsations in a Circular-ribbon Flare}

\author{Q. M. Zhang\altaffilmark{1,2}, D. Li\altaffilmark{1}, and Z. J. Ning\altaffilmark{1}}

\affil{$^1$ Key Laboratory for Dark Matter and Space Science,
Purple Mountain Observatory, CAS, Nanjing 210008, China\email{zhangqm@pmo.ac.cn}}
\affil{$^2$ Key Laboratory of Solar Activity, National Astronomical Observatories, CAS, Beijing 100012, China}

\begin{abstract}
In this paper, we report our multiwavelength observations of the C3.1 circular-ribbon flare SOL2015-10-16T10:20 in active region (AR) 12434. 
The flare consisted of a circular flare ribbon (CFR), an inner flare ribbon (IFR) inside, and a pair of short parallel flare ribbons (PFRs). 
The PFRs located to the north of IFR were most striking in the \textit{Interface Region Imaging Spectrograph} (\textit{IRIS}) 1400 {\AA} and 2796 {\AA} images.
For the first time, we observed the circular-ribbon flare in the Ca {\sc ii} H line of the Solar Optical Telescope (SOT) aboard
\textit{Hinode}, which has similar shape as observed in the Atmospheric Imaging Assembly (AIA) 1600 {\AA} aboard the Solar Dynamic Observatory (\textit{SDO}).
Photospheric line-of-sight magnetograms from the Helioseismic and Magnetic Imager (HMI) aboard \textit{SDO} show that the flare was associated with positive
polarities and a negative polarity inside. The IFR and CFR were cospatial with the negative polarity 
and positive polarities, implying the existence of a magnetic null point (\emph{\textbf{B}}$=$\textbf{0}) and the dome-like spine-fan topology. 
During the impulsive phase of the flare, ``two-step'' raster observations of \textit{IRIS} with a cadence of 6 s and an exposure time of 2 s show plasma 
downflow at the CFR in the Si {\sc iv} $\lambda$1402.77 line ($\log T\approx4.8$), suggesting chromospheric condensation. 
The downflow speeds first increased rapidly from a few km s$^{-1}$ to the peak values of 45$-$52 km s$^{-1}$, before decreasing gradually to the initial levels.
The decay timescales of condensation were 3$-$4 minutes, indicating ongoing magnetic reconnection. Interestingly, the downflow speeds are positively correlated 
with logarithm of the Si {\sc iv} line intensity and time derivative of the \textit{GOES} soft X-ray (SXR) flux in 1$-$8 {\AA}. 
The radio dynamic spectra are characterized by a type \Rmnum{3} radio burst associated with the flare, which implies that the chromospheric condensation was most 
probably driven by nonthermal electrons. Using an analytical expression and the peak Doppler velocity, we derived the lower limit of energy 
flux of the precipitating electrons, i.e., 0.65$\times$10$^{10}$ erg cm$^{-2}$ s$^{-1}$. The Si {\sc iv} line intensity and SXR derivative show quasi-periodic pulsations 
with periods of 32$-$42 s, which are likely caused by intermittent null-point magnetic reconnections modulated by the fast wave propagating 
along the fan surface loops at a phase speed of 950$-$1250 km s$^{-1}$. Periodic accelerations and precipitations of the electrons result in periodic heating observed in the 
Si {\sc iv} line and SXR.
\end{abstract}

Online-only material: animations, color figures

\section{Introduction} \label{s-intro}
Solar flares are one of the most spectacular activities in the solar system. Up to 10$^{29}$$-$10$^{32}$ ergs magnetic free energies are impulsively released, accompanied by rapid 
increases of emissions in various wavelengths \citep{shi11}. It is generally believed that magnetic reconnection plays a key role in the reconfiguration of magnetic field lines and 
conversion of magnetic energy into kinetic and thermal energies of plasma \citep{for96,pri00}. In the context of standard solar flare model \citep{car64,stu66,hir74,kopp76}, the 
accelerated nonthermal electrons propagate downward and heat the chromosphere, forming the bright flare ribbons in H$\alpha$, Ca {\sc ii} H, ultraviolet (UV), and extreme-ultraviolet 
(EUV) wavelengths. The flare ribbons have diverse shapes, with two parallel flare ribbons (PFRs) being the most commonplace \citep{li15c}. Sometimes, circular flare ribbons (CFRs) 
appear accompanied by remote brightenings, which is generally believed to be associated with a magnetic null point (\emph{\textbf{B}}$=$\textbf{0}) and the dome-like spine-fan topology \citep{mas09,zqm12,pon16}. 
More complex flare ribbons have also been observed \citep{jos15}. Apart from the localized heating, hard X-ray (HXR) emissions are generated via Coulomb collisions \citep{bro71}. 
If open magnetic field lines are involved, it is likely that the nonthermal electrons at speeds of 0.06$-$0.25$c$ ($c$ is the speed of light) escape the Sun into the interplanetary space 
and generate type \Rmnum{3} radio bursts with frequency ranging from 0.2 to hundreds of MHz \citep{dul87,asch95,zqm15}. Sometimes, the HXR and radio emissions of a flare show 
quasi-periodic pulsations (QPPs), with their periods ranging from milliseconds \citep{tan10} through a few seconds \citep{kli00,asai01,ning05,naka10,hay16} to several minutes \citep{of06,sych09,naka09,ning14}.
So far, QPPs have been extensively explored using both imaging \citep{su12a,su12b} and spectral observations \citep{mar05}. \citet{li15b} studied the X1.6 flare on 2014 September 10. 
Four-minute QPPs are evident not only in the HXR, EUV, UV, and radio light curves, but also in the temporal evolutions of the Doppler velocities and line widths of the C {\sc i}, O {\sc iv}, 
Si {\sc iv}, and Fe {\sc xxi} lines. \citet{bro15} studied the M7.3 flare on 2014 April 18. The chromospheric and transition region line emissions show quasi-periodic intensity and velocity 
fluctuations with periods of $\sim$3 minutes during the first four peaks and $\sim$1.5 minutes during the last four peaks.

The overpressure of the chromosphere drives evaporation of hot and dense plasma upward into the newly reconnected coronal loops at speeds of tens to hundreds of 
km s$^{-1}$ \citep{fis85a,fis85b,fis85c}. At the same time, downward chromospheric condensation takes place at speeds of a few to tens of km s$^{-1}$ as a result of 
the balance of momentum \citep{ter03}. In the past 40 years, the investigations of chromospheric evaporation and condensation have greatly benefitted from the 
spectroscopic observations in H$\alpha$ and EUV wavelengths from the ground-based as well as space-borne telescopes \citep[e.g.,][]{cza99,bro04,chen10,li11,mil11,you13,tian14,tian15,li15a,pol15,pol16,bat15,gra15}. 
\citet{ich84} studied the red asymmetry of H$\alpha$ line profiles of 4 flares near the disk center. The first two on June 20 and 
June 21 in 1982 were indeed circular-ribbon flares. The red-shifted emission streaks of H$\alpha$ line are caused by the conspicuous downward motion in the flare 
chromospheric region with velocities of 40$-$100 km s$^{-1}$. Using the spectral data of an M6.8 two-ribbon flare observed by the Coronal Diagnostic Spectrometer (CDS) aboard \textit{SOHO}, 
\citet{cza99} found that the Fe {\sc xvi} downward velocity of the regions between the flare ribbons and the magnetic neutral line decreases from $\sim$90 km s$^{-1}$ 
to $\sim$15 km s$^{-1}$ within 2.5 minutes in the gradual phase, indicating ongoing magnetic reconnection. \citet{bro04} studied an M2.0 flare observed by CDS 
and found that the downward velocity of chromospheric condensation in O {\sc v} line decreases from $\sim$40 km s$^{-1}$ to zero within 3 minutes. Compared with CDS, the 
Extreme-ultraviolet Imaging Spectrometer \citep[EIS;][]{cul07} aboard \textit{Hinode} \citep{kos07} covers a much wider temperature range and has much higher resolution and cadence. 
Using the spectral data of a C1.0 flare observed by EIS, \citet{bro13} found that the upward velocities of chromospheric evaporation in Fe {\sc xxiii} line decreases from its maximum value 
($>$200 km s$^{-1}$) to zero within 2 minutes. The successful launch of the \textit{Interface Region Imaging Spectrograph} \citep[\textit{IRIS};][]{dep14} opened a new era of flare 
research, thanks to its unprecedented resolution and cadence. The timescales of chromospheric evaporation range from 5 to 12 minutes in most cases \citep{bat15,li15a,tian15,pol15,pol16,gra15}.
However, it can reach $\sim$20 minutes in an M1.0 flare \citep{sad16}. The timescales of chromospheric condensation range from 1 to 7 minutes \citep{gra15,sad16}. 
A delay of $>$60 s between the starting times of condensation and evaporation has been observed \citep{gra15,you15}.
Both the upward evaporation velocities and downward condensation velocities are positively correlated with the HXR fluxes, which is consistent with the numerical model
of evaporation driven by nonthermal electrons \citep{tian15,li15a}.

So far, chromospheric condensation in circular-ribbon flares has rarely been investigated. Using the imaging and raster observations of a \textit{GOES} C4.2 circular-ribbon 
flare on 2015 October 16 by \textit{IRIS}, \citet{zqm16} found explosive chromospheric evaporation during the impulsive phase of the flare, which is characterized by simultaneous 
plasma upflow (35$-$120 km s$^{-1}$) observed in the high-temperature Fe {\sc xxi} $\lambda$1354.09 line ($\log T\approx7.05$) and downflow (10$-$60 km s$^{-1}$) 
observed in the low-temperature Si {\sc iv} $\lambda$1393.77 line ($\log T\approx4.8$). Based on the quantitative estimation of nonthermal energy flux under the thick-target model and 
the fact that the inner flare ribbon (IFR) where chromospheric evaporation occurred was cospatial with the single HXR source at 12$-$25 keV, the authors concluded that the explosive 
chromospheric evaporation was most likely driven by nonthermal electrons accelerated by magnetic reconnection. However, the \textit{IRIS} observation was in the ``sparse synoptic raster'' 
mode with a cadence of 9.4$\times$36$=$338.4 s, which is difficult to study the temporal evolution of the same position.
There were a couple of homologous circular-ribbon flares in the same active region (AR 12434) on the same day as the C4.2 flare. 
In this paper, we study another one, the C3.1 flare observed by \textit{IRIS}. The raster observation was in the ``two-step raster'' mode with a cadence of 6 s, which is quite suitable for 
investigating the temporal evolution of the same position of flare ribbon. In addition, we observed QPPs during the impulsive phase of the flare.
In Section~\ref{s-data}, we describe the instruments and data analysis using observations from various telescopes. Results and discussions are presented in Section~\ref{s-res} and Section~\ref{s-disc}, respectively. 
Finally, we draw a conclusion in Section~\ref{s-sum}.

\section{Instruments and data analysis} \label{s-data}
\subsection{\textit{SDO} and \textit{Hinode} Observations} \label{s-sdo}
The flare was observed by the Atmospheric Imaging Assembly \citep[AIA;][]{lem12} aboard the \textit{Solar Dynamic Observatory} (\textit{SDO}) in the full-disk images. 
AIA has two UV (1600 {\AA}, 1700 {\AA}) and 7 EUV filters, including 304 {\AA} ($T\approx0.05$ MK), 171 {\AA} ($T\approx0.63$ MK), 193 {\AA} ($T\approx1.5$ MK), 
211 {\AA} ($T\approx2.0$ MK), 335 {\AA} ($T\approx2.5$ MK), 94 {\AA} ($T\approx6.3$ MK), and 131 {\AA} ($T\approx0.37$ MK and 10 MK).
Photospheric line-of-sight (LOS) magnetograms of the AR were observed by the Helioseismic and Magnetic Imager \citep[HMI;][]{sch12} aboard \textit{SDO}. 
The level\_1 data from AIA and HMI were calibrated using the standard Solar SoftWare (SSW) programs \textit{aia\_prep.pro} and \textit{hmi\_prep.pro}, respectively. 
It was also partially observed by the Ca {\sc ii} H ($\sim$3968 {\AA}) filter of the Solar Optical Telescope \citep[SOT;][]{tsu08} aboard \textit{Hinode} with a 
limited field of view (FOV) of $218\arcsec\times109\arcsec$. SOT consists of 4 instruments, including the Broad-band Filter Imager (BFI), Narrow-band Filter Imager 
(NFI), Spectropolarimeter (SP), and Correlation Tracker (CT). BFI produces images in 6 bands (CN bands, Ca {\sc ii} H, G-band, and 3 continuum bands).
The raw BFI data were calibrated using the SSW program \textit{fg\_prep.pro}. The magnetograms and images observed in Ca {\sc ii} H, UV, 
and EUV wavelengths were coaligned with an accuracy of $\sim$0$\farcs$6.

\subsection{\textit{IRIS} Observations} \label{s-iris}
The Slit-Jaw Imager (SJI) aboard \textit{IRIS} observed the flare in Si {\sc iv} $\lambda$1400 ($T\approx0.065$ MK) and Mg {\sc ii} H/K $\lambda$2796 ($T\approx0.01$ MK). 
The raster observation was in the ``two-step raster'' mode (OBS ID: 3660105413). 
As its name implies, each raster had two steps from east to west with a cadence of 6 s and an exposure time of 2 s, 
covering an area of 2$\arcsec$$\times$128$\farcs$42. Each step had a spatial size of 0$\farcs$3327 and a spectral scale of $\sim$25.44 m{\AA} pixel$^{-1}$, which equal to 
$\sim$5.44 km s$^{-1}$ pixel$^{-1}$. The raster data of Si {\sc iv} $\lambda$1402.77 line ($\log T\approx4.8$) were preprocessed using the SSW programs \textit{iris\_orbitvar\_corr\_l2.pro} 
and \textit{iris\_prep\_despike.pro}. Since the intensity of Fe {\sc xxi} $\lambda$1354.09 line was too weak to obtain convincing line profiles, we did not use this line.
The Si {\sc iv} line is an isolated line with red asymmetry. However, the spectra could not been simply fitted by a single-Gaussian or double-Gaussian function.
To determine the line center ($\lambda_c$) and intensity ($I$), we use the same method as described in \citet{bro15}:

\begin{equation} \label{eqn1}
\lambda_c=\frac{\int_{\lambda_1}^{\lambda_2}\lambda\times I_{\lambda}d\lambda}{\int_{\lambda_1}^{\lambda_2}I_{\lambda}d\lambda},
\end{equation}

\begin{equation} \label{eqn2}
I=\int_{\lambda_1}^{\lambda_2}I_{\lambda}d\lambda,
\end{equation}
where $\lambda_1=1402.01$ {\AA} and $\lambda_2=1403.61$ {\AA} signify the lower and upper limits for the integrals. $I_{\lambda}$ denotes the intensity at $\lambda$.

\subsection{Radio Observations} \label{s-radio}
The flare was accompanied by a type \Rmnum{3} radio burst, which was recorded in the radio dynamic spectra by the \textit{WIND}/WAVES \citep{bou95} and two 
ground-based telescopes\footnote{http://e-callisto.org} (\textit{KRIM}, \textit{BLENSW}). WAVES has two detectors: RAD1 (0.02$-$1.04 MHz) and RAD2 (1.075$-$13.825 MHz).
The frequency ranges of \textit{KRIM} and \textit{BLENSW} are 250$-$350 MHz and 10$-$80 MHz, respectively.
The observing parameters of the instruments are summarized in Table~\ref{tbl-1}.

\section{Results} \label{s-res}
\subsection{Flare Ribbons} \label{s-ribb}
\begin{figure}
\epsscale{.80}
\plotone{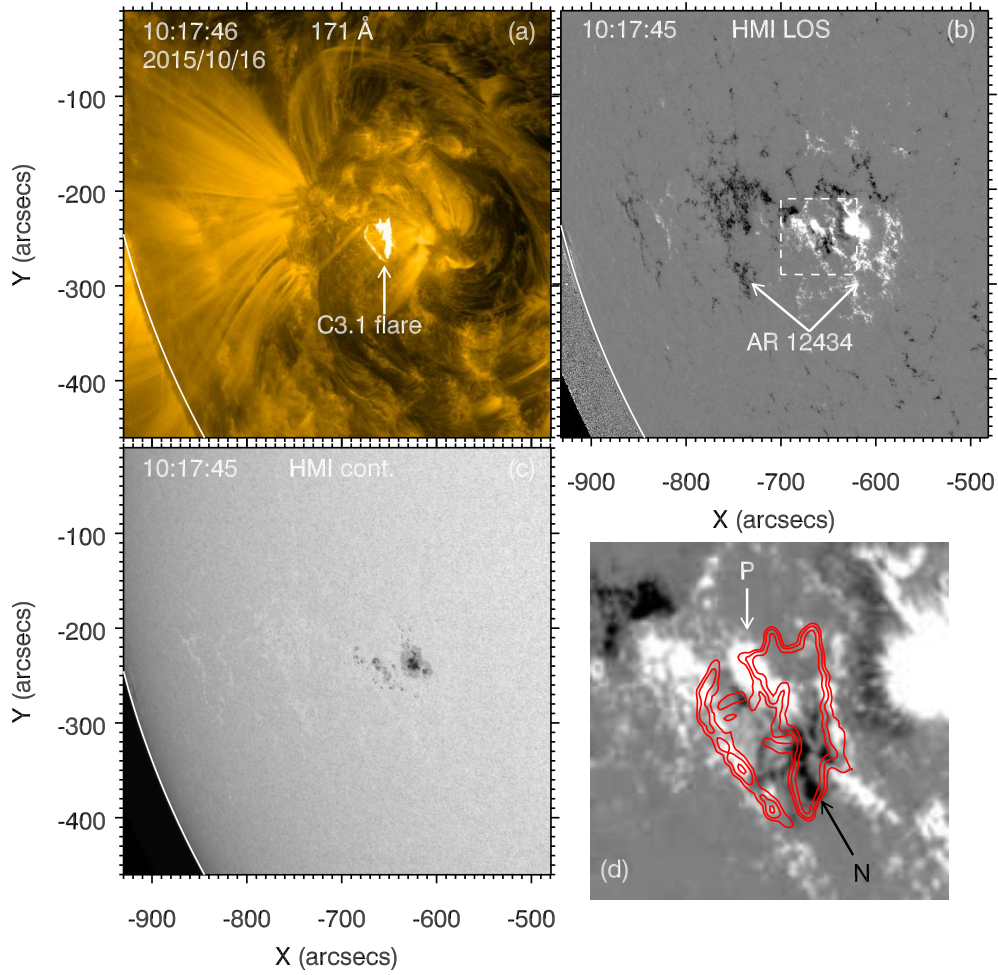}
\caption{(a) AIA 171 {\AA} image at 10:17:46 UT. The white arrow points to the C3.1 cicular-ribbon flare.
(b)-(c) HMI LOS magnetogram and white-light continuum image at 10:17:45 UT. The white arrows point to the main negative and positive polarities of AR 12434.
(d) A close-up of the flare region within the white dashed box of panel (b). The intensity contours of the 171 {\AA} image in panel (a) are overlaid with red lines.
\label{fig1}}
\end{figure}

In Figure~\ref{fig1}, the AIA 171 {\AA} image at 10:17:46 UT is displayed in panel (a). The C3.1 circular-ribbon flare with enhanced emission took place at the center of AR 12434.
Panel (b) shows the HMI LOS magnetogram of the AR at 10:17:45 UT. In order to investigate the magnetic field of the flare, we give a close-up of the region within the dashed 
box in panel (d). As mentioned in \citet{zqm16}, the flare region is characterized by a central negative polarity (N) surrounded by positive polarities (P). The contours of the 171 {\AA} 
intensity in panel (a) are superposed with red lines. It is clear that the flare ribbons were cospatial with the strong magnetic field. Like the C4.2 flare, such a magnetic pattern is
strongly indicative of a magnetic null point and the spine-fan configuration in the corona. The horizontal and vertical length scales of the flare are $L_{x}\approx30\arcsec$ and $L_{y}\approx40\arcsec$, 
respectively. Assuming a semispherical shape of the fan surface, the height of the null point and the loop length ($L$) from the null point to the solar surface are estimated to be 
$0.5\sqrt{L_{x}L_{y}}=17\farcs3$ and $\frac{\pi}{4}\sqrt{L_{x}L_{y}}=27\farcs2$, respectively. The height of null point is higher than that of the coronal bright points on 2007 March 
16 \citep{zqm12}, but slightly lower than that of the M6.7 flare on 2011 September 8 \citep{zqm15}.
Panel (c) demonstrates the HMI white-light continuum image at 10:17:45 UT, which is characterized by the dark sunspot and the surrounding plage of the AR.

\begin{figure}
\epsscale{.80}
\plotone{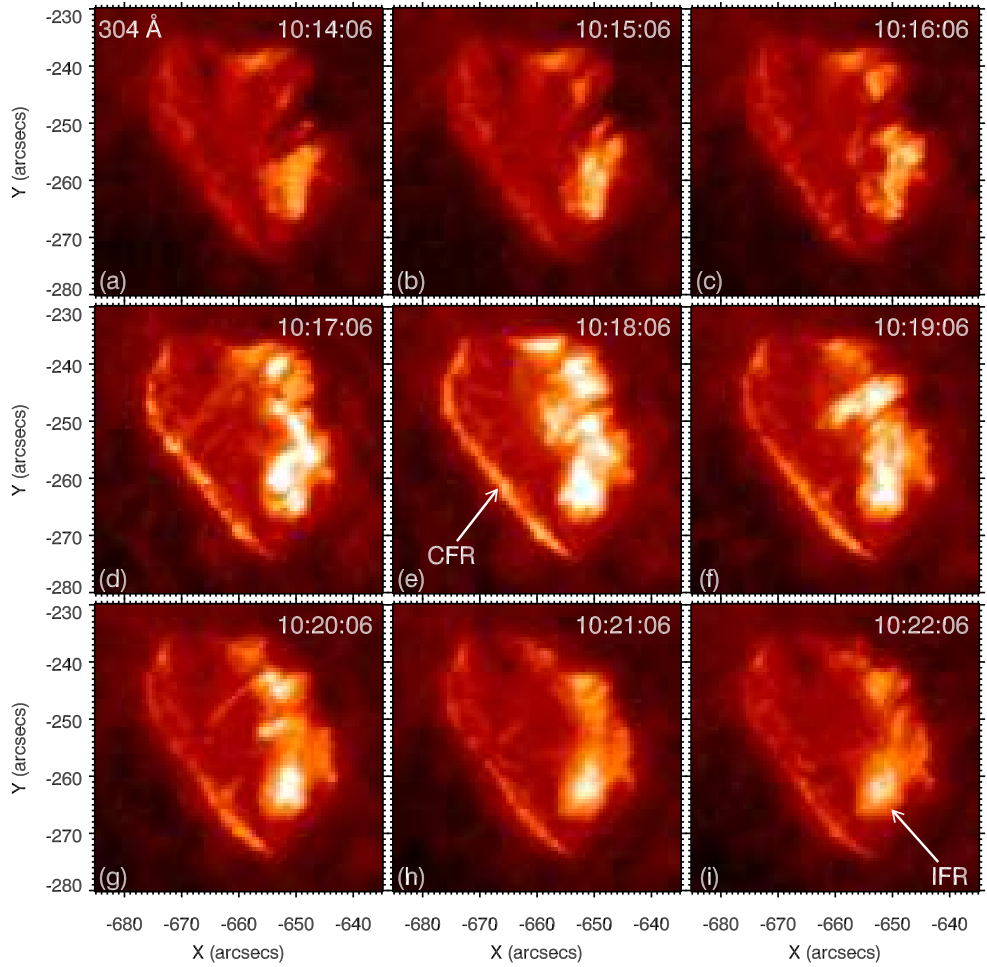}
\caption{Snapshots of the AIA 304 {\AA} images. In panels (e) and (i), the white arrows point to the CFR and IFR, respectively. 
\label{fig2}}
(Animations of this figure are available in the online journal.)
\end{figure}

Figure~\ref{fig2} shows the temporal evolution of the flare in 304 {\AA}. Unlike the C4.2 flare during 13:36$-$13:51 UT, we did not find a preexisting filament residing in the AR 
and its eruption that would result in a blowout jet \citep{zqm16}. The flare started brightening from its southwest part at $\sim$10:15 UT (see panel (b)).
As time goes on, the area of initial brightening expanded and became more complex, including several patches or kernels (see panels (d)-(e) and online movie \textit{anim304.mpg}).
Meanwhile, the narrow eastern part of the flare brightened, forming the CFR as indicated by the arrow in panel (e).
After its peak time around 10:20 UT, the intensity of flare decreased. It is evident that the intensity of CFR decreased much faster than the IFR inside (see panels (g)-(i)).

\begin{figure}
\epsscale{.80}
\plotone{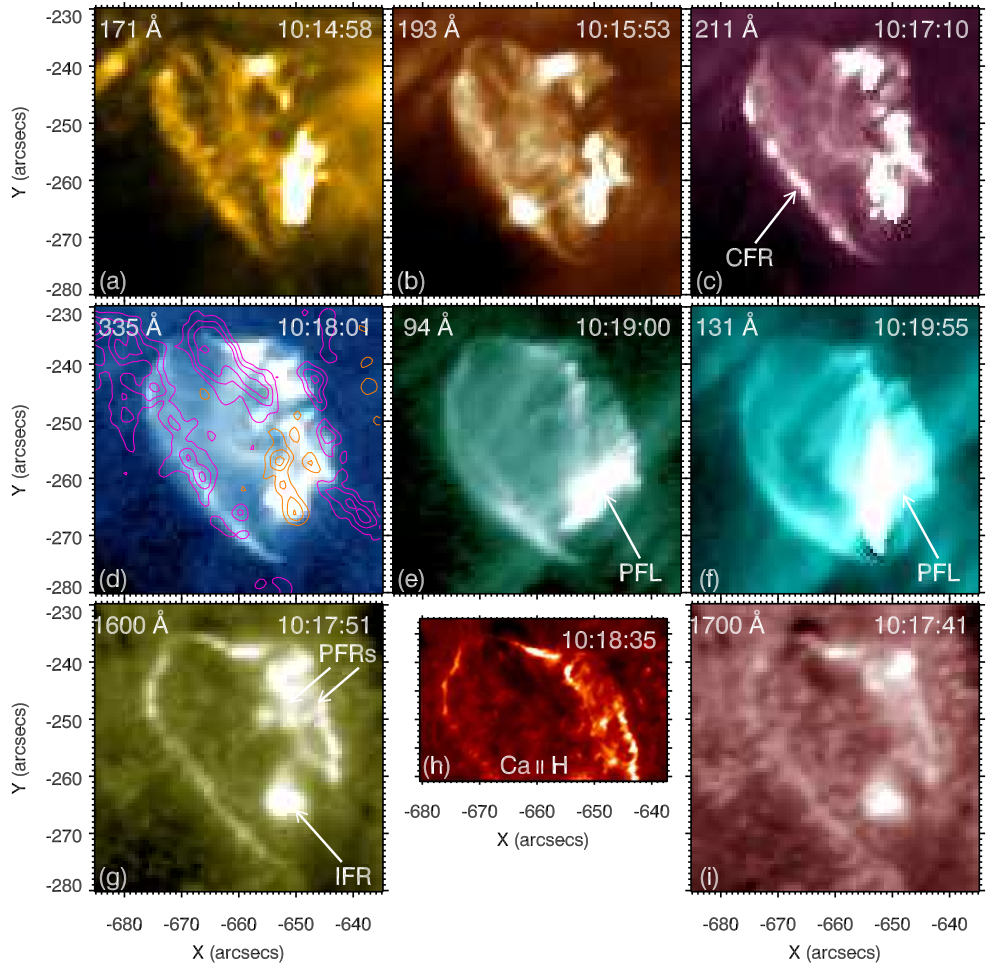}
\caption{Snapshots of the AIA EUV and UV images and SOT Ca {\sc ii} H image.
In panels (c) and (g), the white arrows point to the CFR, IFR, and PFRs. In panel (d), the magenta and orange
lines stand for the contours of the positive and negative LOS magnetic fields, respectively.
In panels (e) and (f), the white arrows point to the bright PFL connecting the PFRs.
\label{fig3}}
(Animations of this figure are available in the online journal.)
\end{figure}

In Figure~\ref{fig3}, the top two rows demonstrate selected images of the flare in the other AIA wavelengths with higher formation temperatures (see online animation \textit{anim335.mpg}). 
The evolution is similar to that in 304 {\AA}. In panel (d), the contours of the positive and negative LOS magnetic fields are overlaid with magenta and orange lines.
Like the C4.2 flare, the CFR was approximately cospatial with positive polarities, while the IFR was approximately cospatial with negative polarity \citep{zqm16}.
The hot, bright post-flare loop (PFL) after being filled with evaporated material was obvious in 94 {\AA} (panel (e)) and 131 {\AA} (panel (f)), which is consistent 
with the peaks around 10:20 UT in the EUV light curves (Figure~\ref{fig6}(a)).
The bottom row shows the CFR and IFR observed in the lower solar atmosphere (1600 {\AA}, Ca {\sc ii} H, and 1700 {\AA}). To our knowledge, this is the first report 
of a circular-ribbon flare observed in Ca {\sc ii} H. The shape of ribbon is similar to that in 1600 {\AA}, but has finer structure owing to the higher resolution of SOT.

\begin{figure}
\epsscale{.80}
\plotone{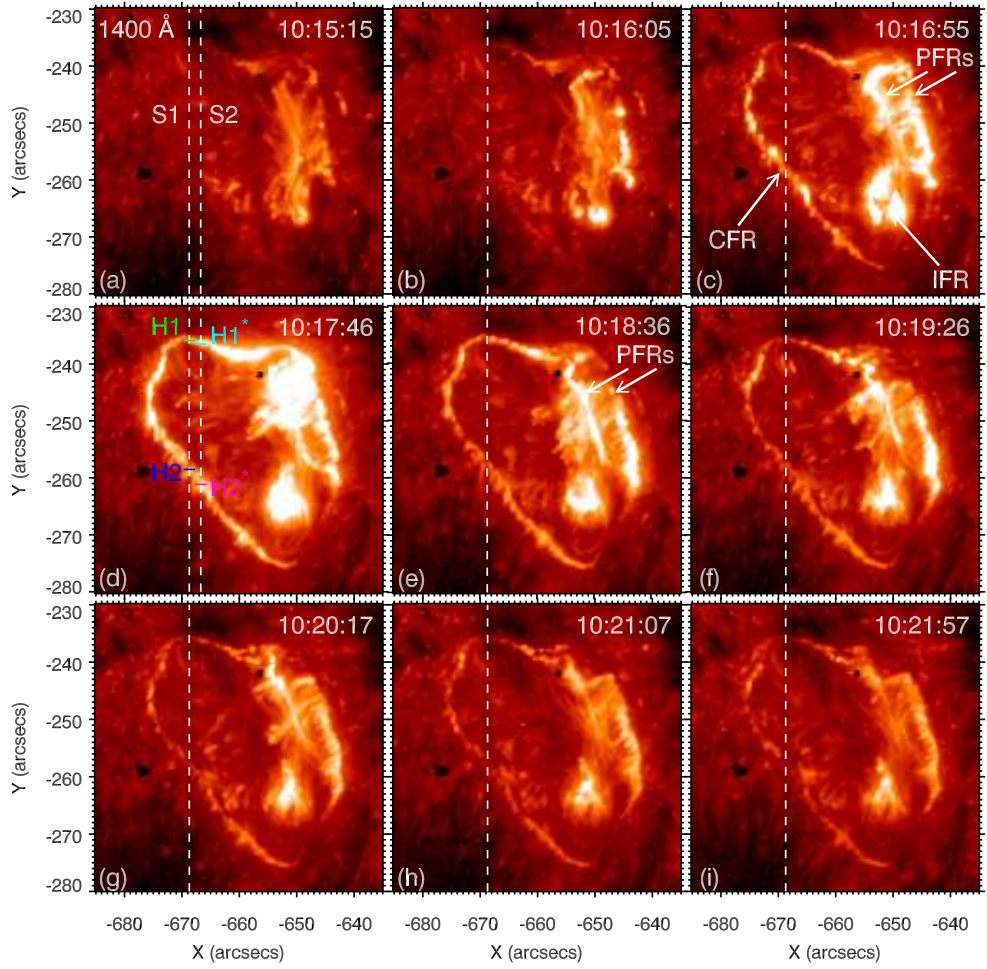}
\caption{Snapshots of the \textit{IRIS}/SJI 1400 {\AA} images. In panel (a), the two vertical dashed lines denote the slit positions of the two-step raster (S1, S2).
In panel (c), the white arrows point to the CFR, IFR, and PFRs. In panel (d), S1 encounters the CFR at H1 and H2. S2 encounters the CFR at H1* and H2*. 
\label{fig4}}
(Animations of this figure are available in the online journal.)
\end{figure}

Figure~\ref{fig4} shows the selected 1400 {\AA} images during 10:15:15$-$10:36:10 UT (see online movie \textit{anim1400.mpg}).
The evolution of the flare is somewhat similar to that in 304 {\AA} in Figure~\ref{fig2}, featuring the bright CFR and IFR with ultrafine structures (see panel (c)).
Despite that the two filters have comparable formation temperatures, new features appear thanks to the high resolution of \textit{IRIS}. 
In panel (c), a pair of bright PFRs are located to the north of IFR. The PFRs are $\sim$18$\farcs$4 in length, which is much shorter than the CFR.
The brightness of PFRs reached the maximum around 10:17:46 UT before decreasing gradually. 
Such a pattern, including the coexisting CFR, IFR, and PFRs, was also observed in the large-scale M7.3 flare on 2014 April 18 \citep{jos15}.
In panel (a), the slit positions of the two-step raster are drawn with vertical dashed lines (S1 and S2). In panel (d), S1 encounters the CFR at H1 and H2. 
S2 encounters the CFR at H1* and H2*. The intensity of H1* is greater than H1, but lower than H2 and H2* in most cases.

\begin{figure}
\epsscale{.80}
\plotone{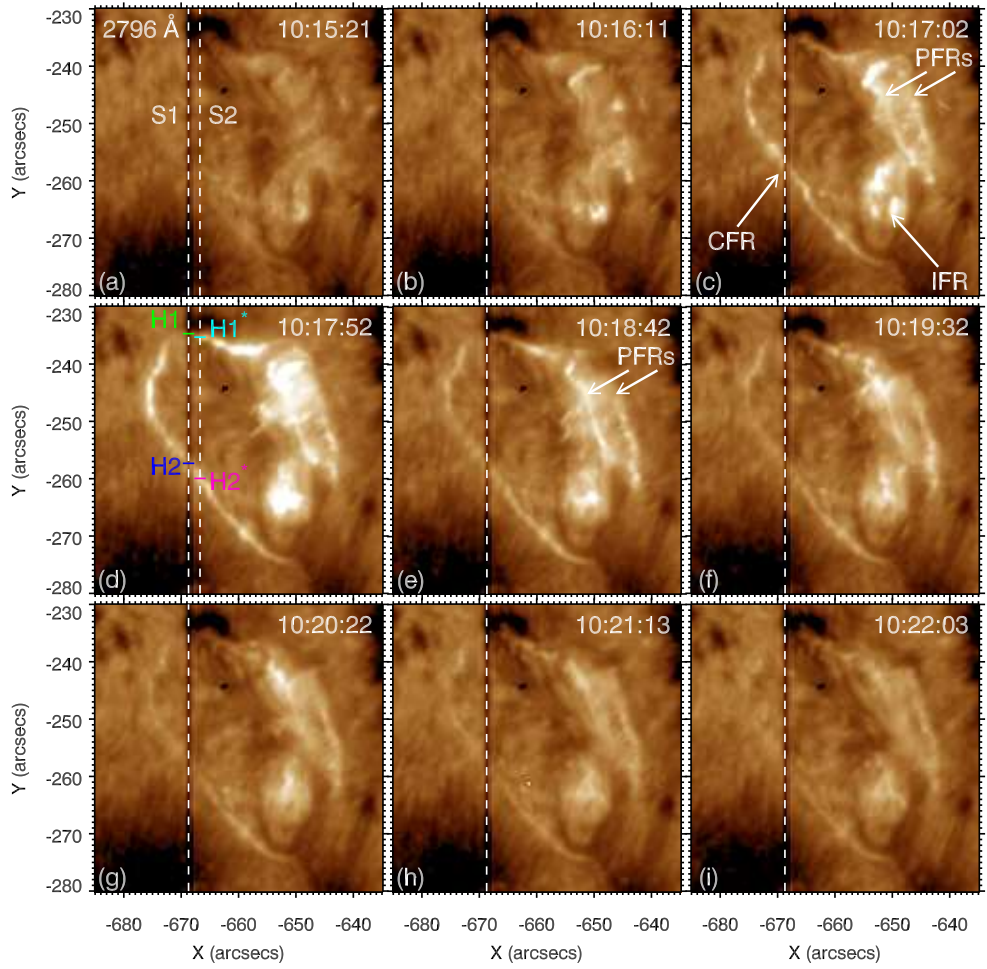}
\caption{Same as Figure~\ref{fig4}, but for \textit{IRIS}/SJI 2796 {\AA}.
\label{fig5}}
\end{figure}

\begin{figure}
\epsscale{.60}
\plotone{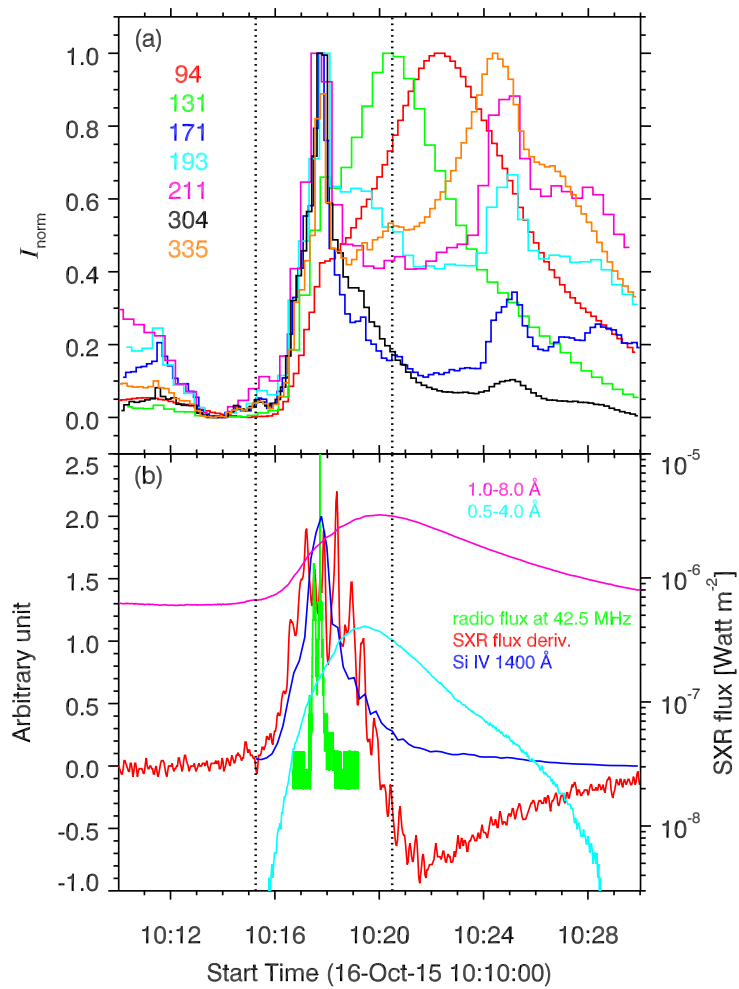}
\caption{(a) Normalized EUV light curves of the flare during 10:10$-$10:30 UT.
(b) SXR light curves of the flare during 10:10$-$10:30 UT in 0.5$-$4 {\AA} (cyan) and 1$-$8 {\AA} (magenta). The time derivative of the 
flux in 1$-$8 {\AA} is drawn with a red line. The light curve of the radio flux at 42.5 MHz is drawn with a green line.
The light curve of the flare in 1400 {\AA} is illustrated with a blue line.
The two dotted lines denote the starting (10:15:15 UT) and ending (10:20:32 UT) times of the \textit{IRIS} raster observation in each panel.
\label{fig6}}
\end{figure}

Figure~\ref{fig5} shows the selected 2796 {\AA} images. The evolution of the flare is similar to that in 1400 {\AA}, featuring the CFR, IFR, and PFRs (see panel (c)).
However, the duration of brightening of the ribbons is much shorter owing to the much lower formation temperature of 2796 {\AA} ($T\approx0.01$ MK).

In Figure~\ref{fig6}, the upper panel shows the normalized light curves of the flare in EUV wavelengths. It is obvious that all the light curves experienced a sharp peak 
around 10:17:46 UT simulantously. For the light curve in 131 {\AA}, the small peak around 10:17:46 UT indicates the existence of hot material up to $\sim$10 MK 
during the early-impulsive phase of the flare \citep{fle13,tian15}. This is also supported by the ``step'' after the fast rise in the 94 {\AA} light curve. As the chromospheric 
evaporation proceeds possibly due to the nonthermal electrons that impact the chromosphere, the coronal loops are filled with hot and dense plasmas, which generate 
strong emissions in the wavelengths with the highest formation temperatures. This is revealed by the dominant peak at $\sim$10:20:10 UT in 131 {\AA} and the following 
peak in 94 {\AA}. As time goes on, the coronal loops cool down primarily due to thermal conduction and create peaks of emissions progressively at decreasing formation 
temperatures of 335 {\AA}, 211 {\AA}, 193 {\AA}, 171 {\AA}, and 304 {\AA}. The cooling timescale of the flare loops from $\sim$10 MK to $\sim$0.6 MK was $\sim$5 minutes.
The lower panel of Figure~\ref{fig6} shows the soft X-ray (SXR) light curves during 10:10$-$10:30 UT in 0.5$-$4 {\AA} (cyan) and 1$-$8 {\AA} (magenta). 
The flare had a short lifetime of $\sim$15 minutes. It started at $\sim$10:15 UT, peaked at $\sim$10:20 UT, and ended at $\sim$10:30 UT.

\subsection{Chromospheric Condensation at the CFR} \label{s-cond}
\begin{figure}
\epsscale{.70}
\plotone{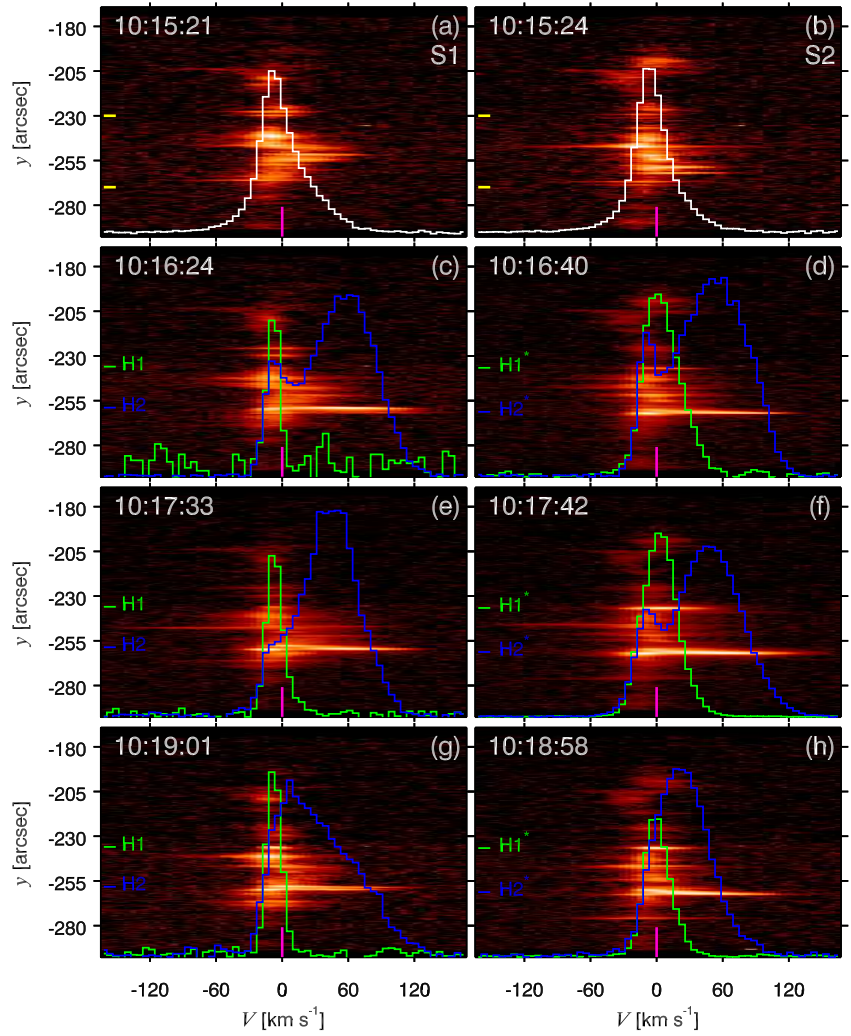}
\caption{\textit{IRIS} spectral window of Si {\sc iv} for S1 (\textit{left panels}) and S2 (\textit{right panels}).
In panels (a) and (b), the white curves represent the spacetime average spectra between the positions (-270$\arcsec$ and -230$\arcsec$) 
marked by the two yellow lines during 10:15:21$-$10:15:43 UT. The rest wavelength ($\lambda_0$) of Si {\sc iv} $\lambda$1402.77 is calculated to be 1402.86$\pm$0.0145 {\AA}.
In panels (c)-(h), the green curves are the spectra at H1 and H1* in three raster scans. The blue curves are the spectra at H2 and H2* in three raster scans.
\label{fig7}}
\end{figure}

In Figure~\ref{fig7}, the \textit{IRIS} spectral window of Si {\sc iv} for S1 and S2 are displayed in the left and right panels, respectively.
In the top two panels, the white curves represent the spacetime average spectra between the vertical positions (-270$\arcsec$ and -230$\arcsec$) marked by the two 
yellow lines during 10:15:21$-$10:15:43 UT. Despite in the very initial phase of the flare, the spectra show weak red asymmetry. Hence, a single-Gaussian fitting is still inapplicable.
Using Equation~(\ref{eqn1}) and the spacetime average spectra, the rest wavelength ($\lambda_0$) of Si {\sc iv} $\lambda$1402.77 is calculated to be 
1402.86$\pm$0.0145 {\AA}, exactly the same value as in \citet{bro15}.
In panels (c)-(h), the green curves are the spectra at H1 and H1* in three raster scans. The blue curves are the spectra at H2 and H2* in three raster scans. 
For H1, the line profiles seem to be slightly blue-shifted, indicating weak plasma upflow. For H1*, the profiles are wider than H1 and show red asymmetry, implying plasma 
downflow. For H2 and H2*, the red asymmetry are much more significant, suggesting faster downflow. The line centers and intensities at each position along S1 and S2
during the raster are calculated using Equations~(\ref{eqn1}) and (\ref{eqn2}). The Doppler velocity ($V_D$) of Si {\sc iv} line is calculated based on the distances 
between line centers and $\lambda_0$:
\begin{equation} \label{eqn3}
V_D=\frac{c}{\lambda_0}(\lambda_{c}-\lambda_{0}).
\end{equation}
According to the uncertainty of $\lambda_0$, the uncertainty of $V_D$ is calculated to be 3.11 km s$^{-1}$.

In Figure~\ref{fig8}, the temporal evolutions of the Si {\sc iv} line intensity and Doppler velocity along slit S1 are demonstrated in the left panels. 
The temporal evolutions of the intensity and Doppler velocity of H1 and H2 are extracted and plotted in the right panels with cyan and magenta lines. 
It is clear that outside the CFR, the intensities and $V_D$ are far less than the CFR. The intensities of H1 and H2 increased from 10:15:15 UT and reached the peak 
values before decreasing with oscillations. The intensity of H2 was greater than H1 in most cases during the flare. The red-shifted $V_D$ of H2 increased quickly 
and reached the maximum of $\sim$45 km s$^{-1}$ at 10:16:24 UT before decreasing gradually to $\sim$2 km s$^{-1}$ at 10:20:29 UT. The timescale of 
chromospheric condensation of H2 was $\sim$4 minutes. The $V_D$ of H1 was blue-shifted and fluctuated slightly around -5 km s$^{-1}$ during most of the lifetime. 
The possible cause of coexisting downflow and upflow along the CFR will be discussed at the end of Section~\ref{s-disc}.

In Figure~\ref{fig9}, the temporal evolutions of the Si {\sc iv} line intensity and Doppler velocity along slit S2 are demonstrated in the left panels.
The temporal evolutions of the intensity and Doppler velocity of H1* and H2* are extracted and plotted in the right panels with cyan and magenta lines. 
The evolutions of intensities of H1* and H2* are similar to H1 and H2. The red-shifted $V_D$ of H2* increased quickly from $\sim$10 km s$^{-1}$ at 10:15:18 UT 
and reached the maximum of $\sim$52 km s$^{-1}$ at 10:17:09 UT before decreasing gradually to $\sim$10 km s$^{-1}$ at 10:20:32 UT. The timescale of 
chromospheric condensation of H2* was $\sim$3.5 minutes. The $V_D$ of H1* was red-shifted and fluctuated slightly around 4 km s$^{-1}$ during most of the lifetime.

As mentioned above, the Si {\sc iv} line intensities and $V_D$ of the CFR have similar trends, implying a temporal correlation between the two parameters.
In Figure~\ref{fig10}, the scatter plots of intensity and $V_D$ are drawn in the left panels for H2 and H2*. Note that the intensities are in $\log$-scale. It is obvious that 
the two parameters are positively correlated with correlation coefficients $\geq$0.8.

\subsection{QPPs of the Flare} \label{s-qpp}
In Figure~\ref{fig6}(b), the time derivative of SXR flux in 1$-$8 {\AA} is plotted with a red line, which is characterized by several regular peaks in the impulsive phase. 
Using the same method as described in \citet{li15b}, we decompose the SXR derivative into a slow-varying component plus a rapid-varying component, the later of 
which is displayed in Figure~\ref{fig11}(a). It is clearly revealed that the SXR derivative fluctuated quasi-periodically. The Morlet wavelet transform of the fast-varying 
component is illustrated in Figure~\ref{fig11}(b). The quasi-periodicity, i.e., QPPs during 10:16:20$-$10:21:10 UT, is confirmed. The periods are 25$-$40 s with its maximum 
power lying at $\sim$32 s.

\begin{figure}
\epsscale{.80}
\plotone{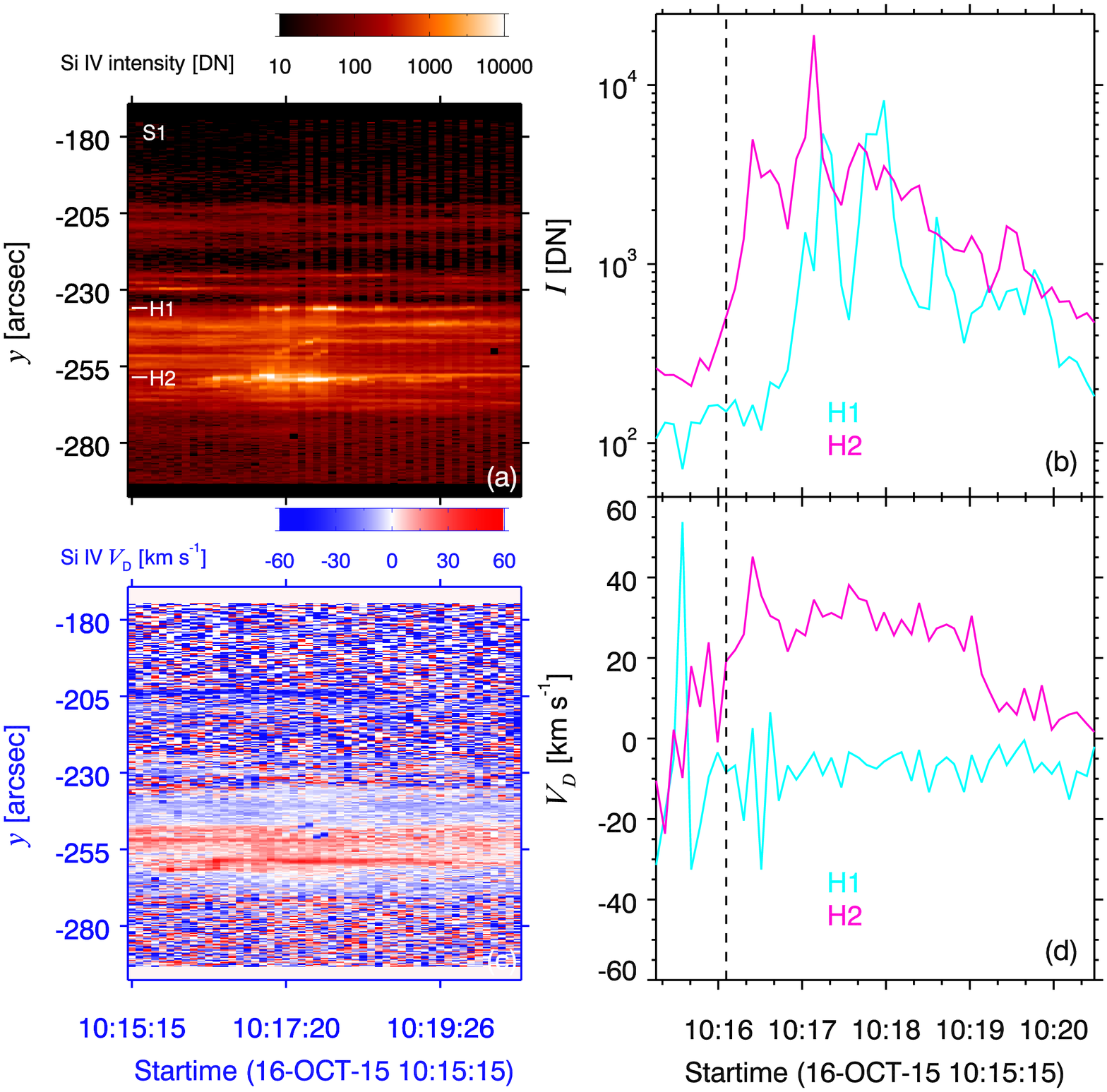}
\caption{(a, c) Temporal evolutions of the Si {\sc iv} line intensity and Doppler velocity along slit S1. 
(b, d) Temporal evolutions of the intensity and Doppler velocity of H1 and H2. The dashed line denotes the starting time for performing correlation analysis in Figure~\ref{fig10}.
\label{fig8}}
\end{figure}

\begin{figure}
\epsscale{.80}
\plotone{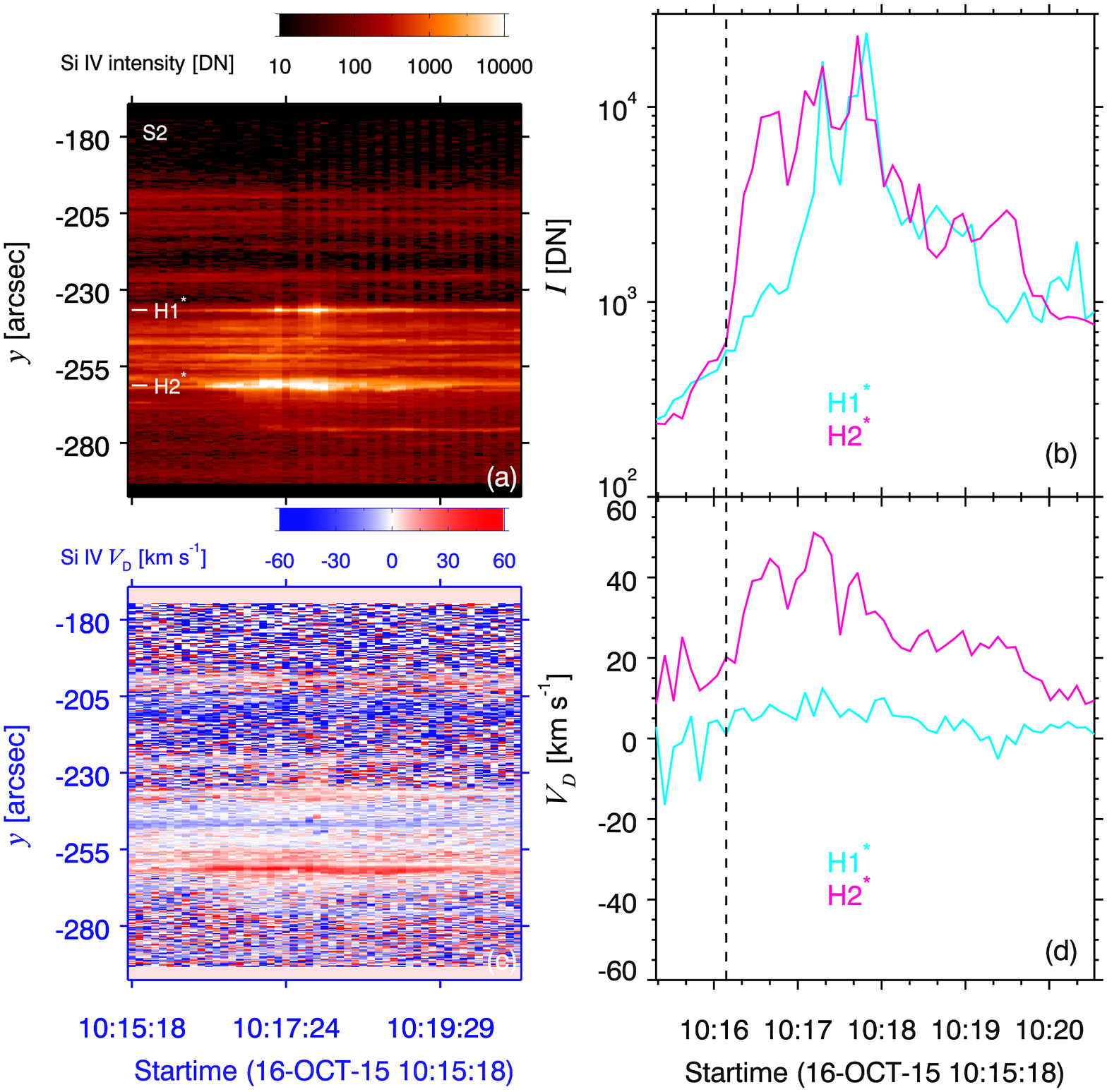}
\caption{(a, c) Temporal evolutions of the Si {\sc iv} line intensity and Doppler velocity along slit S2. 
(b, d) Temporal evolutions of the intensity and Doppler velocity of H1* and H2*. The dashed line denotes the starting time for performing correlation analysis in Figure~\ref{fig10}.
\label{fig9}}
\end{figure}

In Figure~\ref{fig8} and Figure~\ref{fig9}, the violent fluctuations of the Si {\sc iv} intensities of the CFR imply the existence of QPPs in UV wavelengths. In order to justify the conjecture, 
we decompose the light curves in Figure~\ref{fig8}(b) and Figure~\ref{fig9}(b) into slow-varying components and fast-varying components by smoothing. The Morlet wavelet transforms 
of the fast-varying components are illustrated in Figure~\ref{fig12}. The QPPs in UV wavelengths during 10:16:05$-$10:19:05 UT are confirmed. 
In the left panels, the periods range from 25 to 60 s, with the maximum power lying at $\sim$42 s for H1 and $\sim$38 s for H2. 
In the right panels, the periods range from 25 to 50 s, with the maximum power lying at $\sim$35 s for H1* and $\sim$33 s for H2*.
It is evident that the periods are very close and slightly longer than the period in SXR (32 s). 
The evolutions of $V_D$ in Figure~\ref{fig8}(d) and Figure~\ref{fig9}(d), however, do not show QPPs. 
In Figure~\ref{fig6}(b), the light curve of the flare in 1400 {\AA} is drawn with a blue line. By and large, the total 1400 {\AA} intensity, as a signature of heating at the 
chromosphere by nonthermal electrons, has similar evolution to the SXR derivative. Correlation analysis shows that the two parameters are temporally correlated 
(c.c.$\approx$0.77) during the impulsive phase. However, QPP was absent in the 1400 {\AA} light curve.

\section{Discussion} \label{s-disc}
\subsection{What Is the Cause of Chromospheric Condensation?} \label{s-cause1}
The observational studies of chromospheric evaporation and condensation have a long history. \citet{ich84} studied the temporal evolution of the H$\alpha$ line profiles of 4 flares 
near the disk center. The timescales of the total evolutions are $\sim$1 minute. The authors proposed that the red asymmetry of H$\alpha$ line results from the downward motion 
of the compressed chromospheric region produced by the impulsive heating by energetic electron beam or thermal conduction. 
\citet{li15a} found positive correlations between the HXR emissions and upward Doppler shifts of Fe {\sc xxi} line and downward Doppler shifts of C {\sc i} line ($\log T\approx4.0$) 
in two X1.6 flares. In the C4.2 circular-ribbon flare, \citet{zqm16} found that the IFR where explosive chromospheric evaporation occurred was cospatial with the single HXR source. 
A quantitative estimation of the energy flux of the nonthermal electron supports the electron-driven evaporation/condensation. 

\begin{figure}
\epsscale{.80}
\plotone{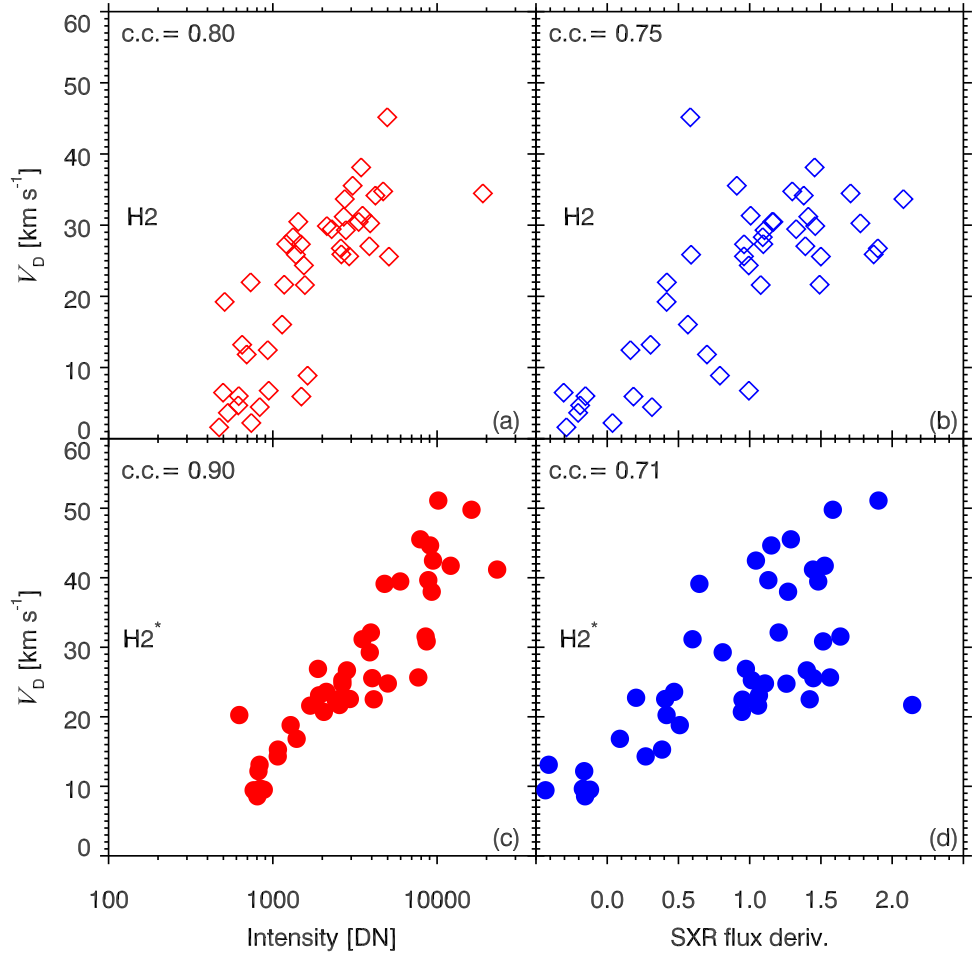}
\caption{Scatter plots of Si {\sc iv} Doppler velocity and intensity (\textit{left panels}) and  
SXR flux deriviative (\textit{right panels}) for H2 (\textit{top panels}) and H2* (\textit{bottom panels}).
Note the intensities are in $\log$-scale.
The correlation coefficients are also displayed in each panel.
\label{fig10}}
\end{figure}

\begin{figure}
\epsscale{.60}
\plotone{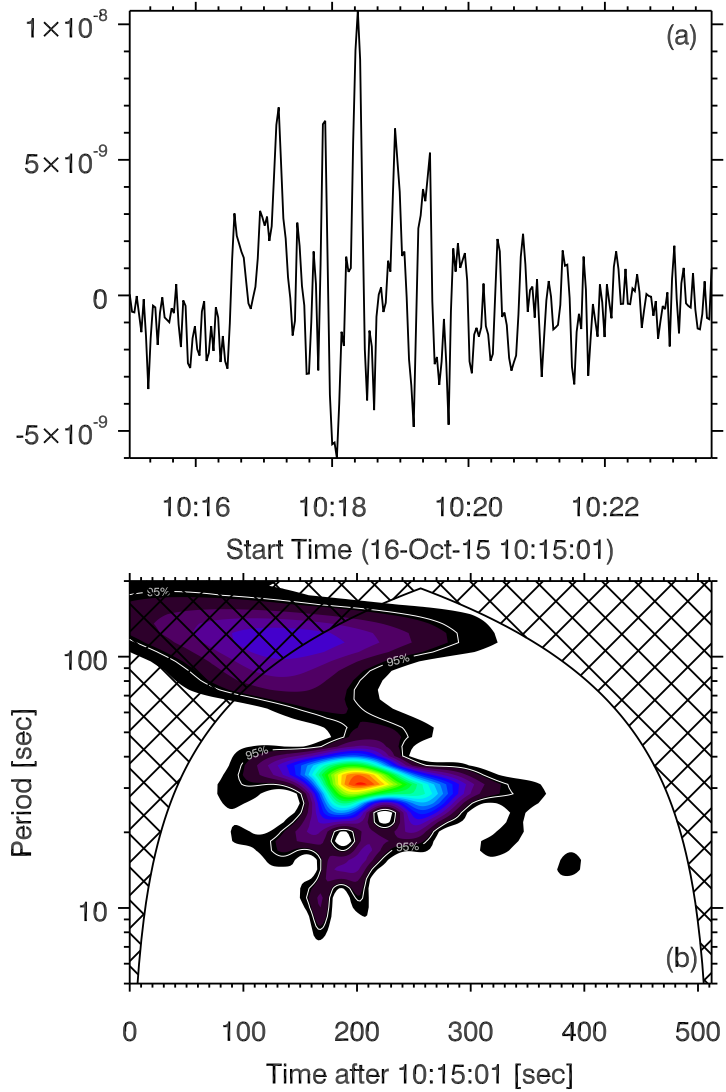}
\caption{(a) Fast-varying component of the time derivative of SXR flux in Figure~\ref{fig6}(b). (b) Morlet wavelet transform of the component. 
The red color represents the highest power. The white, solid contour encloses regions of $\geq$95\% confidence level for a white noise 
process \citep{tor98}. The cross-hatched regions indicate the ``cone of influence", where edge effects become important \citep{wang09}.
\label{fig11}}
\end{figure}

\begin{figure}
\epsscale{.80}
\plotone{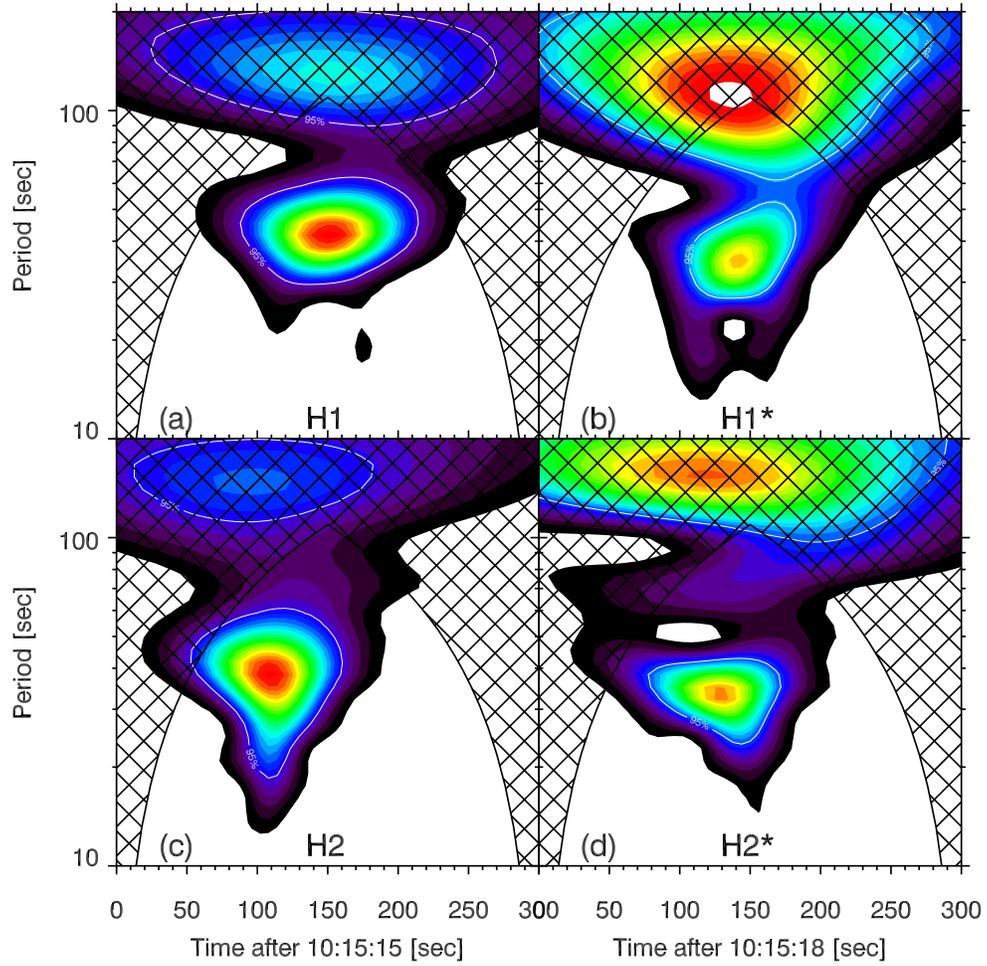}
\caption{Morlet wavelet transforms of the fast-varying components of the Si {\sc iv} intensities of H1 (a), H1* (b), H2 (c), and H2* (d).
The annotations have the same meanings as in Figure~\ref{fig11}(b).
\label{fig12}}
\end{figure}

\begin{figure}
\epsscale{.80}
\plotone{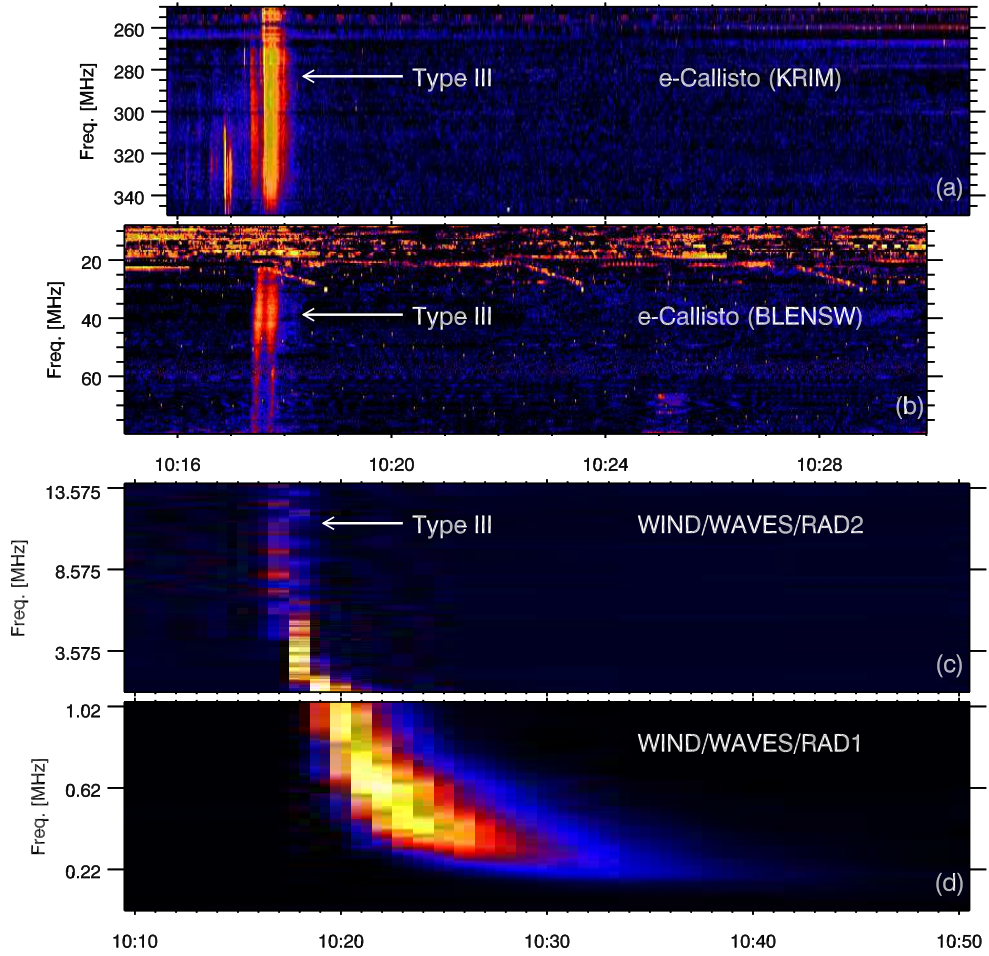}
\caption{Radio dynamic spectra recorded by \textit{KRIM} (a), \textit{BLENSW} (b), and \textit{WIND}/WAVES (c-d). 
The time is in UT and the frequency is in MHz. The arrows point to the type \Rmnum{3} radio burst with enhanced emissions and fast frequency drift rate.
\label{fig13}}
\end{figure}

Like in \citet{li15a}, we explored the relationship between the downward $V_D$ of the CFR and SXR derivative for the C3.1 flare. 
The scatter plots of the two parameters are drawn in the right panels of Figure~\ref{fig10} for H2 and H2*, indicating that they are positively correlated with correlation coefficients $\geq$0.7.
In Figure~\ref{fig13}, the radio dynamic spectra from \textit{KRIM} and \textit{BLENSW} are displayed in the top two panels. 
The most striking feature is the type \Rmnum{3} burst with enhanced emissions and rapid frequency drift during 
10:17:00$-$10:18:00 UT. The dynamic spectra from RAD1 and RAD2 aboard \textit{WIND}/WAVES are displayed in the bottom two panels. The frequency of burst drifted rapidly from 
13 MHz at 10:18 UT to 1 MHz at 10:20 UT and then drifted slowly to $\sim$0.2 MHz until $\sim$10:40 UT. The radio flux at 42.5 MHz is plotted in Figure~\ref{fig6}(b) with a green line.
The peaks of the radio flux were roughly coincident with the peaks of SXR derivative. Considering that the HXR flux and type \Rmnum{3} radio burst origin from the nonthermal electrons propagating downward 
into the chromosphere along the newly reconnected magnetic field lines and upward into the interplanetary space along the open field lines, the coincidence of the peaks during the impulsive phase 
of the flare implies the existence of nonthermal electrons accelerated by magnetic reconnection. Combining the positive correlation between $V_D$ of the CFR and SXR derivative, we conclude 
that the chromospheric condensation of the C3.1 flare was most probably driven by nonthermal electrons.

The investigations of chromospheric condensation in theory and numerical simulations have also made huge progress. \citet{fis86} derived an analytical expression of the condensation lifetime:
\begin{equation} \label{eqn4}
\tau=\pi\sqrt{H/g},
\end{equation}
where $H$ and $g$ stand for the gravitational scale height and acceleration of gravity in the chromosphere. The typical value of $\tau$ is $\sim$1 minute.
However, superposition of continuous condensations energized at different times within an unresolved observational element naturally result in a longer condensation lifetime \citep{fis89}.
In the combined modeling of acceleration, transport, and hydrodynamic response in solar flares, \citet{rub15} found that the condensation timescale can reach 70$-$90 s.
In the 1D radiative hydrodynamic numerical simulations, \citet{reep15} found that the timescales of the electron-driven chromospheric evaporation are 3.5$-$4 minutes for different cases.

In Table~\ref{tbl-2}, we list the timescales of chromospheric condensation and evaporation in the previous literatures. It is revealed that the timescales of evaporation upflows 
(2$-$20 minutes) are generally 2$-$3 times larger than the timescales of condensation downflows (1$-$7 minutes). For the C3.1 flare in this study, the condensation timescale 
is 3$-$4 minutes, which is consistent with previous findings. To our knowledge, the cadence (6 s) of the \textit{IRIS} raster observation is the highest ever for the 
investigation of chromospheric condensation.

Based on the radiative hydrodynamic flare model, \citet{fis89} derived an analytical expression of the peak downflow speed for explosive evaporation:
\begin{equation} \label{eqn5}
v_{peak}\approx0.6(F_{evap}/\rho_{ch})^{1/3},
\end{equation}
where $F_{evap}$ and $\rho_{ch}$ represent the energy flux and chromospheric density. For the C3.1 flare, $v_{peak}=52$ km s$^{-1}$. Taking 10$^{-11}$ g cm$^{-3}$ as the typical 
value of $\rho_{ch}$ in the chromosphere, $F_{evap}$ is estimated to be 0.65$\times$10$^{10}$ erg cm$^{-2}$ s$^{-1}$, which is close to the threshold energy flux for explosive
evaporation ($\sim$10$^{10}$ erg cm$^{-2}$ s$^{-1}$). The estimated $F_{evap}$ should be a lower limit since the slit positions (S1 and S2) are located at 
the CFR whose maximum intensity is far less than the IFR and PFRs. According to the correlation between the intensity and $V_D$ of Si {\sc iv} line in the left panels of Figure~\ref{fig10}, 
the peak downward velocities at the IFR and PFRs should be larger and the nonthermal electron energy flux should be higher.

\subsection{What Is the Cause of QPPs?} \label{s-cause2}
QPPs of solar flares have been extensively observed and reported in various wavelengths. However, the mechanism of QPPs still remains unclear. 
On one hand, there are diverse MHD waves (kink mode and sausage mode) in the solar atmosphere, such as the slow wave, fast wave, and Alfv\'{e}n wave.
The existence and propagation of waves may modulate the emissions in certain wavelengths \citep[e.g.][]{naka04,mar05,tan10,su12a}. On the other hand, 
the magnetic reconnection rate may be modulated by the $p-$mode wave \citep{chen06}, slow-mode wave \citep{naka11}, Alfv\'{e}n wave \citep{asai01}, 
fast sausage-mode wave \citep{zai82}, and fast kink-mode wave \citep{naka06}. 
Therefore, the acceleration and propagation of electrons in the downward and upward directions would be modulated by the periodic magnetic reconnections \citep{asch94,ning05}. 
Besides, the tearing-mode instability and coalescence of magnetic islands occur iteratively, resulting in intermittent magnetic reconnection and particle acceleration \citep{kli00,kar04}.
In a single loop flare on 2005 January 1, \citet{naka10} found QPPs of $\gamma$-ray emission with a period of $\sim$40 s, which are also present in the HXR and microwave 
emissions. They proposed that the QPPs are created by periodic magnetic reconnections accompanying particle acceleration triggered by a global kink oscillation in a nearby coronal loop.
In an M1.8 flare on 2002 October 20, \citet{zim10} found QPPs of both thermal and nonthermal HXR emissions with periods of 16 s and 36 s, which are interpreted in terms of MHD 
oscillations excited in two interacting systems of flaring loops. 

In our study, the QPPs with periods of 32$-$42 s are observed in UV (Si {\sc iv} $\lambda$1402.77) and SXR derivative.
As described at the beginning of Section~\ref{s-ribb}, the length of a loop from the null point to the solar surface is estimated to be $\sim$20 Mm.
If the QPPs are caused by the modulation of magnetic reconnection by fast wave propagating along the fan surface, the phase speed ($2L/P_{QPP}$) is estimated to be 950$-$1250 km s$^{-1}$, 
which is close to the typical value for fast wave in the corona. Hence, the QPPs in the C3.1 flare are most probably caused by the modulation of magnetic reconnection, particle 
acceleration, and subsequent precipitation by fast wave propagating along the fan surface loops. The nonthermal electrons propagating downward heat the chromosphere 
and generate evaporation as well as condensation. Therefore, QPPs are generated in both UV and SXR during the impulsive phase of the flare.
In the X1.6 two-ribbon flare on 2014 September 10, QPPs are observed in multiwavelengths from radio to HXR \citep{li15b}. Our results indicate that QPPs exist not only in the typical 
two-ribbon flares, but also in circular-ribbon flares. The same mechanism may at work in both types.

Finally, we briefly discuss the influence of the rest wavelength on the results. As mentioned in Section~\ref{s-iris}, the spectral profiles of Si {\sc iv} $\lambda$1402.77 line could not been fitted 
by a single-Gaussian function or double-Gaussian function. Using the same method as described in \citet{bro15}, we derived the same rest wavelength of Si {\sc iv} line, i.e., 1402.86$\pm$0.0145 {\AA}.
Therefore, our method is reasonable and results are convincing. The blueshifted Doppler velocities of H1 (see Figure~\ref{fig8}(d)) may be a signature of gentle chromospheric evaporation 
due to a very low energy flux of electrons. However, the blueshifts are too small to draw a solid conclusion.

\section{Summary} \label{s-sum}
In this paper, we report our multiwavelength observations of the C3.1 circular-ribbon flare by various instruments on 2015 October 16. The main results are summarized as follows:
\begin{enumerate}
\item{The flare brightened first at the southwest part. As time goes on, the area of initial brightening expanded and became more complex, including several patches or kernels. 
Meanwhile, the narrow eastern part of flare appeared, forming the CFR that encircled the IFR. Apart from the CFR and IFR, there were a pair of short PFRs to the north of IFR, 
which were most striking in the \textit{IRIS}/SJI 1400 {\AA} and 2796 {\AA}. For the first time, we observe the circular-ribbon flare in the Ca {\sc ii} H line aboard 
\textit{Hinode}/SOT. The shape of CFR is similar to that observed in AIA 1600 {\AA}.}
\item{Like the C4.2 homologous flare, this flare was associated with positive polarities and a negative polarity inside in the AR 12434.
The IFR and CFR were cospatial with the negative polarity and positive polarities, implying the existence of a magnetic null point and dome-like spine-fan topology.}
\item{High-cadence (6 s) \textit{IRIS} raster observations show plasma downflow in the Si {\sc iv} line during the impulsive phase of the flare. 
The Doppler velocities increased from a few km s$^{-1}$ to the peak values of 45$-$52 km s$^{-1}$. Afterwards, the speeds decreased gradually to the initial levels. 
The timescales of the chromospheric condensation were 3$-$4 minutes, indicating ongoing magnetic reconnection.}
\item{The Doppler velocities of the condensation are positively correlated with the logarithm of the Si {\sc iv} line intensities and time derivative of the SXR flux in 1$-$8 {\AA}.
The radio dynamic spectra reveal that the flare was accompanied by a type \Rmnum{3} radio burst. Combining the correlations and radio burst, we conclude that the 
chromospheric condensation was most probably driven by nonthermal electrons. Using the analytical expression in \citet{fis89} and the peak downflow speed, we carried out 
a diagnostic of energy flux of the precipitated electrons. A lower limit of the flux is estimated to be 0.65$\times$10$^{10}$ erg cm$^{-2}$ s$^{-1}$.}
\item{During the impulsive phase of the flare, the Si {\sc iv} line intensities and SXR derivative show QPPs with periods of 32$-$42 s. To our knowledge, this is the first 
report of QPPs in circular-ribbon flares. The QPPs were most probably caused by intermittent
null-point magnetic reconnections modulated by the fast wave propagating along the fan surface loops at a phase speed of 950$-$1250 km s$^{-1}$. 
The intermittent acceleration and precipitations of electrons result in periodic heating observed in the Si {\sc iv} line and emissions in SXR.}
\end{enumerate}

In the future, additional case studies using state-of-the-art, multiwavelength observations and MHD numerical simulations are required to figure out the causes of chromospheric 
condensation and QPPs in circular-ribbon flares.

\acknowledgements
The authors appreciate the referee for inspiring and valuable comments. We also acknowledge H. S. Ji, Y. N. Su, M. D. Ding, and L. Feng for fruitful discussions.
\textit{IRIS} is a NASA small explorer mission developed and operated by LMSAL with mission operations executed at NASA Ames Research Center 
and major contributions to downlink communications funded by the Norwegian Space Center (NSC, Norway) through an ESA PRODEX contract.
\textit{SDO} is a mission of NASA\rq{}s Living With a Star Program. AIA and HMI data are courtesy of the NASA/\textit{SDO} science teams. 
\textit{Hinode} is a Japanese Mission, with NASA and STFC (UK) as international partners. 
This work is supported by NSFC (Nos. 11303101, 11333009, 11573072, 11603077), the Fund of Jiangsu Province (Nos. BK20161618, BK20161095),
and the open research program of Key Laboratory of Solar Activity, National Astronomical Observatories, CAS (No. KLSA201510).

\clearpage

\begin{table}
\caption{Description of the observational parameters.}
\label{tbl-1}
\centering
\begin{tabular}{l c c c c}
\hline\hline
Instrument & $\lambda$ & Time & Cadence & Pixel Size \\
           &     ({\AA})   & (UT)  &    (s)    &  (arcsecs)  \\
\hline
  \textit{SDO}/AIA & 94$-$335 & 10:10:00$-$10:30:00 & 12 & 0.6 \\
  \textit{SDO}/AIA & 1600, 1700 & 10:10:00$-$10:30:00 & 24 & 0.6 \\
  \textit{SDO}/HMI &  6173      & 10:10:00$-$10:30:00 & 45 & 0.5 \\
  \textit{IRIS}/SJI & 1400, 2796 & 10:15:15$-$10:36:10 & 12.6 & 0.333 \\
  \textit{IRIS}/raster & Si {\sc iv} 1402.77 & 10:15:15$-$10:20:32 & 6 & 0.333 \\
  \textit{Hinode}/SOT & Ca {\sc ii} H 3968 & 10:00:33$-$10:38:33 & 120 & 0.218 \\
  \textit{GOES}   & 0.5$-$4, 1$-$8  &  10:10:00$-$10:50:00 & 2.04  & $-$ \\
  \textit{WIND}/WAVES & 0.02$-$13.825 MHz & 10:10:00$-$10:50:00 & 60 & $-$ \\
  \textit{KRIM} & 250$-$350 MHz & 10:16:00$-$10:31:00 & 0.25 & $-$ \\
  \textit{BLENSW} & 10$-$80 MHz & 10:15:00$-$10:30:00 & 0.25 & $-$ \\
\hline
\end{tabular}
\end{table}

\clearpage

\begin{table}
\caption{List of the timescales ($\tau$ in minute) of chromospheric condensation and evaporation in previous literatures.}
\label{tbl-2}
\centering
\begin{tabular}{l c c c c}
\hline\hline
Author & Flare class & Date & Type & $\tau$ \\
\hline
\citet{ich84} & 1B & 1982 & condensation & $\leq$1 \\
\citet{cza99} & M6.8 & 1998/04/29 & condensation & $\sim$2.5 \\
\citet{bro04} & M2.0 & 2001/04/24 & condensation & $\sim$3 \\
\citet{gra15} & X1.6 & 2014/09/10 & condensation & $\sim$1 \\
\citet{sad16} & M1.0 & 2014/06/12 & condensation & $\sim$7 \\
\hline
\citet{bro13} & C1.0 & 2012/03/07 & evaporation & $\sim$2 \\
\citet{gra15} & X1.6 & 2014/09/10 & evaporation & $\sim$12 \\
\citet{bat15} & X1.0 & 2014/03/29 & evaporation & $\sim$5 \\
\citet{pol15} & C6.5 & 2014/02/03 & evaporation & $\sim$6 \\
\citet{pol16} & X2.0 & 2014/10/27 & evaporation & $\sim$13 \\
\citet{sad16} & M1.0 & 2014/06/12 & evaporation & $\sim$20 \\
\hline
\end{tabular}
\end{table}

\clearpage

\clearpage

\clearpage

\clearpage

\clearpage

\clearpage

\clearpage

\clearpage

\clearpage

\clearpage

\clearpage

\clearpage

\clearpage

\end{document}